\documentclass[
showpacs,
floatfix,
aps,
prb,
amsmath,
twocolumn,
groupaddress,
]{revtex4}
\usepackage{amsmath,amssymb}
\usepackage{amscd, latexsym}
\usepackage{mathrsfs}
\usepackage{graphicx} 
\usepackage{epstopdf} 
\usepackage{cancel} 
\usepackage{amsfonts}
\usepackage{exscale}
\usepackage{dcolumn}
\usepackage{bm}
\usepackage{color}

\usepackage{lmodern}
\usepackage[T1]{fontenc}
\usepackage[latin9]{inputenc}
\usepackage{verbatim}
\usepackage{float}

\begin{document}

\title{Quantum Photovoltaic Effect in Double Quantum Dots}

\author{Canran Xu}
\author{Maxim G. Vavilov}
\affiliation{Department of Physics, University of Wisconsin-Madison, Wisconsin 53706, USA }

\date{January 16, 2013}

\begin{abstract}
We analyze the photovoltaic current through a double quantum dot system coupled to a high-quality driven microwave resonator.  The conversion of photons in the resonator to electronic excitations produces a current flow even at zero bias across the leads of the double quantum dot system.  We demonstrate that due to the quantum nature of the electromagnetic field in the resonator, the photovoltaic current exhibits a  double peak dependence on the frequency $\omega$ of an external microwave source.
The distance between the peaks is determined by the strength of interaction between photons in the resonator and electrons in the double quantum dot. The double peak structure disappears as strengths of relaxation processes increases, recovering a simple classical condition for maximal current when the microwave frequency is equal to the resonator frequency.

\end{abstract}


\maketitle

\section{Introduction}

The interaction of electrons in conductors with
electromagnetic fields has long been considered within a classical picture
of electromagnetic (EM) radiation.
A widely--known example is the photon assisted tunneling (PAT) in double quantum dot (DQD) systems,\cite{VanderWiel2002} when the EM field brings an electron trapped 
at the ground state to an excited state and facilitates electron transfer. 
This classical description of the EM field breaks in high-quality microwave resonators based on superconducting transmission line geometry.\cite{Wallraff2004} 
Interaction of such EM fields with electronic devices require a quantum treatment known as the circuit quantum electrodynamics (cQED).\cite{Blais04,You2011}

Recently, several experimental groups studied systems consisting of a superconducting high quality resonator and a DQD\cite{Frey2012a,Delbecq2011,Toida2012,Chorley2012,Petersson2012} or a voltage biased Cooper pair box.\cite{Astafiev2007}
The coupling strength between a resonator photon mode and electron states
in a DQD is characterized by the vacuum Rabi frequency $g$ with reported values
in the range of $g/2\pi \sim 10^8$Hz.
These systems call for re-examination of the PAT by taking into account a quantum description of the EM field in terms of photon excitations. One may expect at least 
two important distinctions from the classical treatment:
(1) the Lamb shift that renormalizes quantum states of electrons and photons; 
(2) spontaneous photon emission that breaks symmetry between absorption and emission processes and is important in systems with either a finite voltage bias between the leads\cite{Childress2004,Sanchez2007,Jin2011} or an inhomogeneous temperature distribution.\cite{Vavilov2006}

\begin{figure}
\centering{}\includegraphics[width=0.97\columnwidth]{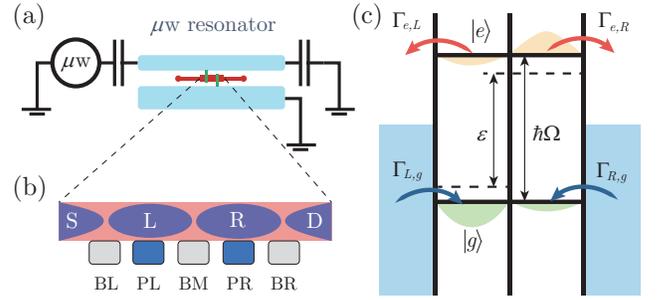}
\caption{\label{fig:The-DQD-configuration}
(Color online) (a) An illustration of a DQD and a transmission line resonator coupled to an external microwave source $\mu$w. 
(b) A schematic view of the DQD.  Electrons are confined to the left (L) and right (R) dots by barrier gates BL, BM, and BR that also control electron tunneling rates between the source, S, and the left dot, the left and right dots, and the right dot and the drain, D, respectively. 
Electrostatic energies of two quantum dots are defined by the plunger gates, PL and PR,  and the PL gate is also connected to an antinode of the resonator, see \textit{e.g.} Refs.~\onlinecite{Petersson2012,Toida2012}. 
(c) Electronic states of the DQD are presented in  both the eigenstate basis (solid lines) and the left--right basis (dashed lines). 
Tunneling from the excited state, $|e\rangle$, to the left/right lead, with rate $\Gamma_{e,L/R}$ and from the left/right lead to the ground state, $|g\rangle$, with rate $\Gamma_{L/R,g}$ are illustrated by arrows. 
}
\end{figure}

In this paper we study the photovoltaic current through a DQD coupled to a high-quality microwave resonator at zero bias across the DQD.
The resonator is driven by external microwave source that populates a photon mode of the resonator, see Fig.~\ref{fig:The-DQD-configuration}.  The photons excite electrons in the DQD and produce electric current even at zero bias, similar to the classical PAT case.\cite{VanderWiel2002,Stoof1996,Pedersen1998} 
We show that due to the coupling of electrons and photons, the current as a function of the source frequency has a multiple peak structure with splitting between the peaks determined by the coupling strength $g$ and reflects the Lamb shift of electronic energy states. 
We also demonstrate that the interaction-induced splitting is sensitive to the energy and phase relaxation rates in the DQD.

We note that the photovoltaic effect discussed here is a common phenomenon when 
the current in an electronic circuit is generated by out-of-equilibrium EM environment. Examples of this phenomenon include 
the current response of a DQD in the vicinity of the biased
quantum point contact\cite{Ouyang2010} or another circuit element  out of equilibrium\cite{Silva2007} with the electronic system.  However, 
because out-of-equilibrium photons of the environment have a broad spectrum, the generated current does not exhibit a resonant dependence on parameters of the system that we observe in a system of a single mode high quality resonator and a DQD.

\section{Model}

We consider a DQD system with each dot connected to its individual
electron reservoir at zero temperature and at zero bias between the reservoirs, see Fig.~\ref{fig:The-DQD-configuration}(b,c).
The gate voltages of the DQD are adjusted near a triple point of its
stability diagram.\cite{VanderWiel2002} To be specific, we choose
a triple degeneracy point between $(N_{l},N_{r})$, $(N_{l}+1,N_{r})$
and $(N_{l},N_{r}+1)$ electron states in the DQD and denote these
states as $\left|0\right\rangle $, $\left|L\right\rangle $ and $\left|R\right\rangle $,
respectively. We model the system by the Hamiltonian $\tilde{H}=H_{\text{DQD}}+H_r+H_{\text{int}}$, where $H_{\text{DQD}}$ describes states with an extra electron in
the left or right dot, $\left|L\right\rangle $ or $\left|R\right\rangle $:
\begin{equation}
H_{\text{DQD}}=\frac{1}{2}\varepsilon\tau_{z}+\mathcal{T}\tau_{x},\label{eq:1}
\end{equation}
with $\varepsilon$ being the electrostatic energy difference between
the two states, and $\mathcal{T}$ being the tunnel matrix element of an
electron between the dots. The Pauli matrices are defined in the
subspace of states $\left|L\right\rangle $ and $\left|R\right\rangle $
as $\tau_{x}=\left|R\right\rangle \left\langle L\right|+\left|L\right\rangle \left\langle R\right|$
and $\tau_{z}=\left|R\right\rangle \left\langle R\right|-\left|L\right\rangle \left\langle L\right|$.
A resonator driven by an external microwave
source  is described by the Hamiltonian
\begin{equation}
H_{r}=\hbar\omega_{0}a^{\dagger}a+2\hbar {{F}}(a^{\dagger}+a)\cos\omega t \label{eq:1res}
\end{equation}
with $a$ $(a^{\dagger})$ denoting the annihilation (creation) operators
for microwave photons in the resonator, $\hbar {{F}}$ being
the amplitude of the external drive of the resonator and $\omega_{0}$ ($\omega$) being   
frequency of the resonator (source).  
The interaction
between the microwave field and the DQD system is represented by\cite{Childress2004}
\begin{equation}
H_{\text{\text{int}}}=\hbar g_{0}(a^{\dagger}+a)\tau_{z}.\label{eq:2}
\end{equation}
This interaction describes the shift of energy difference
between states $\left|R\right\rangle $ and $\left|L\right\rangle $
due to the electric potential of the plunger gates defined by the microwave photon field. We assume that the photon field is distributed between the left and right plunger gates, see Fig.~\ref{fig:The-DQD-configuration}(b) and does not influence the source and drain voltage to avoid the rectification effects.\cite{Brouwer2001,DiCarlo2003,Vavilov2005d}

Further calculations are more convenient in the basis of the ground,
$\left|g\right\rangle $, and excited, $\left|e\right\rangle $
states of the Hamiltonian, Eq.(\ref{eq:1}):
\begin{align}
\left|e\right\rangle  & =\cos(\theta/2)\left|L\right\rangle +\sin(\theta/2)\left|R\right\rangle ,\nonumber \\
\left|g\right\rangle  & =-\sin(\theta/2)\left|L\right\rangle +\cos(\theta/2)\left|R\right\rangle .\label{eq:4}
\end{align}
Here $\theta=\arctan(2\mathcal{T}/\varepsilon)$ characterizes the
hybridization of the $\left|L\right\rangle $ or $\left|R\right\rangle $ states. The energy splitting between the eigenstates $\hbar\Omega=\sqrt{\varepsilon^{2}+4\mathcal{T}^{2}}$
can be tuned  independently by varying $\varepsilon$ and $\mathcal{T}$ via dc gate voltages.
We further eliminate the time-dependence in Hamiltonian
Eq.(\ref{eq:1res}) by applying unitary operator ${\cal U}=\exp(-i\omega t(a^{\dagger}a+\sigma_{z}/2))$
and utilize the rotating frame approximation to obtain\cite{Childress2004, Jin2011} 
\begin{align}
\frac{{H}}{\hbar}  =& \frac{1}{\hbar}{\cal U}^{\dagger}\tilde{H}{\cal U}-i\frac{\partial{\cal U}^{\dagger}}{\partial t}{\cal U}=\frac{\Omega-\omega}{2}\sigma_{z} \label{eq:5-1}\\
 & + (\omega_{0}-\omega)a^{\dagger}a+g(a\sigma^{+}+a^{\dagger}\sigma^{-})+F(a^{\dagger}+a),\nonumber 
\end{align}
where  $g=g_{0}\sin\theta$ characterizes
the actual strength of the coupling between the microwave field and DQD
states responsible for photon absorption or emission, the Pauli
matrices $\sigma_{z}=\left|e\right\rangle \left\langle e\right|-\left|g\right\rangle \left\langle g\right|$, $\sigma^{-}=\left|g\right\rangle \left\langle e\right|$ and $\sigma^{+}=\left|e\right\rangle \left\langle g\right|$
are defined in terms of eigenstates of the electron Hamiltonian,
Eq. (\ref{eq:1}).

We analyze the behavior of the system with Hamiltonian Eq. (\ref{eq:5-1})
in the presence of relaxation in electron and photon degrees of freedom
by employing the Born-Markov master equation for the full density
matrix 
\begin{equation}
\dot{\rho}={\cal L}_{\text{tot}}\rho=-\frac{i}{\hbar}\left[H,\rho\right]+{\cal L}_{\text{diss}}\rho.\label{eq:drho}
\end{equation}
The first term on the r.h.s. of Eq.(\ref{eq:drho}) describes the unitary evolution
of the system and the second term accounts for the dissipative processes
in the resonator and DQD systems\cite{Ouyang2010}
\begin{equation}
\begin{split}
&{\cal L}_{\text{diss}}\rho  \equiv\kappa{\cal D}(a)\rho+\gamma{\cal D}(\sigma^{-})\rho+\frac{\gamma_{\phi}}{2}{\cal D}(\sigma_{z})\rho\label{eq:Lrho}
\\
 & +(\Gamma_{L,g}+\Gamma_{R,g}){\cal D}(c_{g}^{\dagger})\rho+(\Gamma_{e,L}+\Gamma_{e,R}){\cal D}(c_{e})\rho,
\end{split}
\end{equation}
where ${\cal D}(x)\rho=\left(2x\rho x^{\dagger}-x^{\dagger}x\rho-\rho x^{\dagger}x\right)/2$
is the Lindblad superoperator. The relaxation of the photon field
in the resonator with rate $\kappa$ is represented by $\kappa{\cal D}(a)\rho$ and the electron relaxation from the excited state $\left|e\right\rangle $ to the ground state $\left|g\right\rangle $ with rate $\gamma$ is
represented by $\gamma{\cal D}(\sigma^{-})\rho$. The last two Lindblad superoperators account for the loading of the ground state $\left|g\right\rangle $ and unloading of the excited state $\left|e\right\rangle $
of the double quantum dot via electron tunneling in
terms of operators $c_{e}=\left|0\right\rangle \left\langle e\right|$
and $c_{g}^{\dagger}=\left|g\right\rangle \left\langle 0\right|$,
respectively. The tunneling rates 
$\Gamma_{L,g}=\Gamma_{l}\cos^{2}(\theta/2)$, $\Gamma_{R,g}=\Gamma_{r}\sin^{2}(\theta/2)$,  $\Gamma_{e,L}=\Gamma_{l}\sin^{2}(\theta/2)$ and $\Gamma_{e,R}=\Gamma_{r}\cos^{2}(\theta/2)$ are written in terms of tunneling rates $\Gamma_{l/r}$ in the basis of $\left|L\right\rangle $ and $\left|R\right\rangle $ states.

\begin{figure}
\begin{centering}
\includegraphics[width=0.95\columnwidth]{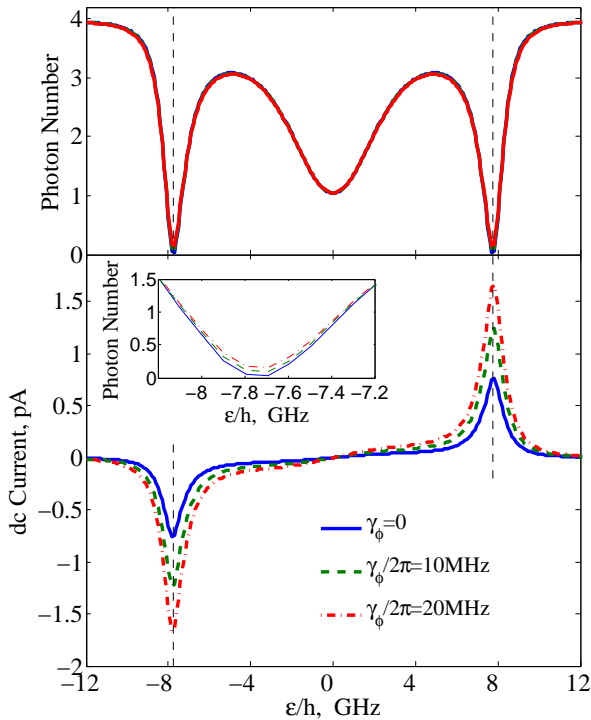}
\end{centering}
\caption{\label{fig:2}(Color online) The average number of photons in the resonator and the photovoltaic
current as functions of level bias $\varepsilon$ for ${\cal{T}}/2\pi=1$
GHz, ${{F}}=50\text{ } \mu\text{s}^{-1}$ and $\omega_{0}/2\pi=8\text{ GHz}$. The current
is generated near the resonant condition when $\varepsilon =\pm\sqrt{\hbar^2\omega_0^2-4{\cal{T}}^2}$ (vertical lines).
The three curves represent different dephasing rates $\gamma_\phi$ the DQD.}
\end{figure}

Note that in Eq.(\ref{eq:drho}), the dynamics of state $\left|0\right\rangle $
only appears via the tunneling terms involving $D(c_{e})\rho$ and
${\cal D}(c_{g}^{\dagger})\rho$. These terms can be categorized by
whether the empty state is loaded from the left or right lead with
coefficients depending on projection of the eigenstates onto the left/right
states, as shown in Fig. \ref{fig:The-DQD-configuration}. In this
picture,\cite{Ouyang2010} the photovoltaic current 
is given by
\begin{equation}
I=e\Gamma_{r}\left(\cos^{2}\frac{\theta}{2}\left\langle e\right|\bar{\rho}\left|e\right\rangle -\sin^{2}\frac{\theta}{2}\left\langle 0\right|\bar{\rho}\left|0\right\rangle \right).\label{eq:8}
\end{equation}
in terms of the reduced density matrix $\bar{\rho}=\mathrm{Tr}_{\rm{ph}}\{\rho\}$,
where we traced out photon degrees of freedom of the resonator.
We also analyze the number of photons in the resonator, 
\begin{equation}
\bar{N}=\text{Tr}\left\{a^{\dagger}a\rho\right\},
\label{eq:N}
\end{equation} 
where we trace out both photon and electron degrees of freedom.

\begin{figure}
\begin{center}
\includegraphics[width=0.95\columnwidth]{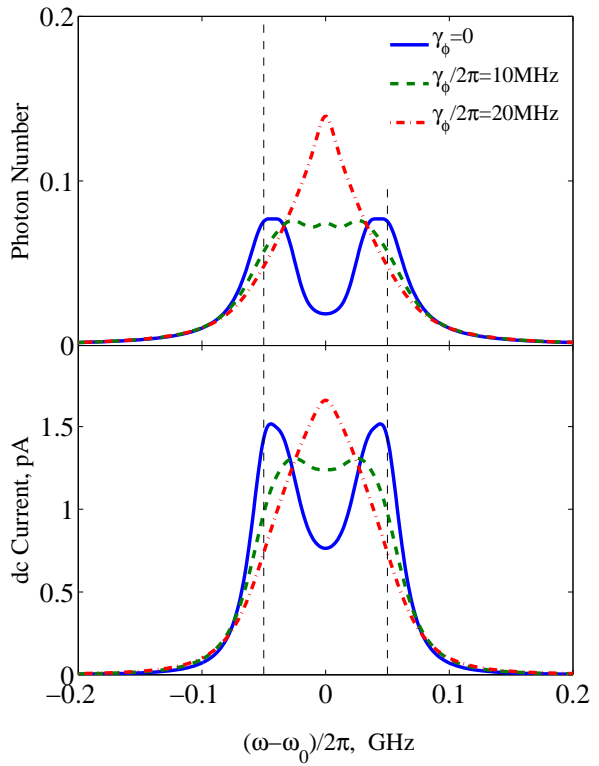}
\end{center}
\caption{\label{fig:3}(Color online) The average number of photons in the resonator and the
photovoltaic current as a function of the frequency $\omega$ of the microwave
drive for ${\cal{T}}/2\pi=1$ GHz, ${{F}}=50\text{ $\mu$s}^{-1}$ and $\Omega=\omega_{0}=2\pi\times 8\text{ GHz}$, $g/2\pi = 48.5\text{ MHz}$.
For $\gamma_\phi=0$, both average number of photons $\bar N$ and the photovoltaic current show local minima at $\omega=\omega_0$ and local maxima near $\omega=(E_{1,\pm}-E_0)/\hbar$, shown by vertical lines. As the  dephasing rate $\gamma_\phi$ increases, the double peaks merge to a single peak at $\omega=\omega_0$.
}
\end{figure}

\section{Results}

The average number of photons in the resonator, $\bar N$, and the dc component of photocurrent can be found using the steady state solution of the master equation, \eqref{eq:drho}, with $\dot \rho=0$.  We numerically find the full density matrix $\rho$ for a double quantum dot and photon field of the resonator in the Fock's space using Quantum Optics Toolbox\cite{Tan1999} and QuTiP\cite{Johansson2012}, both of which provide consistent results.  The steady state solution for the density matrix $\rho$ defines the average number of photons $\bar N$, Eq. \eqref{eq:N}, and the photocurrent, Eq.\eqref{eq:8}. 

Our choice of parameters is motivated by Ref.~[\onlinecite{Toida2012}].  We choose the  relaxation rate $\gamma=2\pi\times25\text{ MHz}$, the resonator relaxation rate $\kappa/2\pi=8\text{ MHz}$, tunneling amplitude between the individual dots ${\cal{T}}/2\pi=1\text{ GHz}$, the tunneling rate from a dot to a lead $\Gamma_{l/r}=2\pi\times 30\text{ MHz}$, and the resonator frequency $\omega_0/2\pi=8$GHz.  
We note that to keep the coupling constant finite, we have to take ${\cal{T}}\sim \hbar\Omega$, since  $g=g_0\sin\theta$, Eq.\eqref{eq:5-1}, vanishes for ${\cal{T}}=0$. Below we fix $g_0/2\pi=\text{ }200$MHz.

First, we investigate dependence of the photocurrent on
the separation between energy levels in the double quantum dot, controlled by the electrostatic energy difference $\varepsilon$. We take frequency  $\omega$  of microwave source to be equal to the resonator frequency, $\omega=\omega_0$, and fix the drive amplitude ${{F}}=50\text{ $\mu$s}^{-1}$.
Dependence of the average number of photons in the resonator and the photocurrent on energy $\varepsilon$ is presented in Fig.~\ref{fig:2} for three values of the dephasing rate $\gamma_\phi/2\pi= 0,\ 10,\ 20$MHz. 
As the energy difference between the excited and ground states of the quantum dot goes through the resonance  $\Omega=\omega_{0}$, we observe a significant suppression of the photon number in the resonator,  
see the top panel and the inset in Fig.~\ref{fig:2}.
This is expected behavior because the DQD system enhances photon absorption in
the resonator at $\Omega\simeq\omega_0$.  Absorbed photons cause transitions between the ground
and excited electronic states.  These electrons tunnel to the leads and generate electric current though the DQD.
This current is shown in the lower panel of Fig.~\ref{fig:2} and is peaked at $\varepsilon =\pm\sqrt{\hbar^2\omega_0^2-4{\cal{T}}^2}$ or $\varepsilon/(2\pi \hbar)\simeq \pm7.75$GHz, indicated by dashed vertical lines.

\begin{figure}
\begin{center}
\includegraphics[width=0.95\columnwidth]{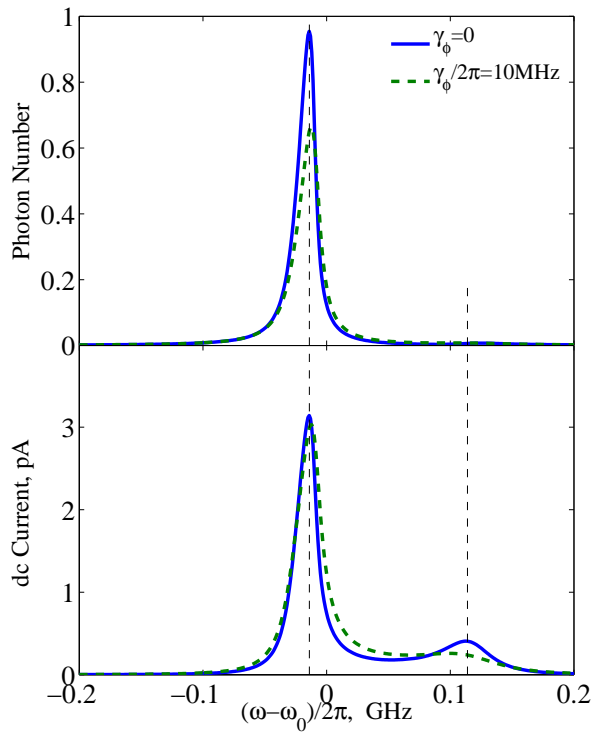}
\par\end{center}
\caption{\label{fig:4}(Color online) The average number of photons in the resonator and the
photovoltaic current as a function of the frequency $\omega$ of the microwave
drive for detuned DQD and resonator system with $\Omega/2\pi=8.1\text{ GHz}$,  $\omega_{0}/2\pi=8.0\text{ GHz}$,  the intradot tunneling  ${\cal{T}}/2\pi=1$ GHz, and the drive amplitude  ${{F}}=50$ MHz. 
The photon average number has a peak
at $\omega=(E_{1,-}-E_0)/\hbar$, Eq.~\eqref{eq:JC}, while the photovoltaic
current exhibits a double peak feature at $\omega=(E_{1,\pm}-E_0)/\hbar$ (vertical lines). }
\end{figure}

One feature in Fig.~\ref{fig:2} is that the photon number is also reduced
at zero bias $\varepsilon$, when the photovoltaic current vanishes.
This suppression is a result of strong enhancement of the coupling constant $g=g_{0}$  at $\varepsilon=0$, resulting in stronger dissipation in the resonator and increase of off-resonant absorption rate. At the same time the photovoltaic current
vanishes at $\varepsilon=0$ due to cancellation between  the two terms in Eq.(\ref{eq:8}).

The curves for the photon number and the current do  not significantly change after the dephasing rate $\gamma_\phi$ is introduced in addition to the energy relaxation rate $\gamma$.  
Dephasing smears the resonant condition for the photon absorption by the DQD and has two effects: 
(1) the number of photons increases a little near the resonance $\Omega\simeq \omega_0$, see the inset in Fig.~\ref{fig:2};   
(2) the resonant absorption of photons by electrons is suppressed  resulting in reduction of the photocurrent.  
We note that in the case presented in Fig.~\ref{fig:2} the first effect is stronger than the second effect and dephasing increases the magnitude of photocurrent for the case of fixed $\omega=\omega_0$.

Next, we consider the case when the frequency of the microwave source,  $\omega$, is varied
while the energy splitting $\hbar\Omega $ of the DQD and the resonator frequency $\omega_0$ are fixed.  The microwave radiation is mostly reflected when its frequency does not match the difference between energies $E_{n,\pm}$ of the resonator and  DQD system defined by the Jaynes-Cummings spectrum:
\begin{equation}
E_{n,\pm}=n\hbar\omega_0\pm \frac{\hbar}{2}\sqrt{4g^2n+\Delta^2}, \quad E_{0}=\frac{\hbar\Delta}{2},\label{eq:JC}
\end{equation} 
where $\Delta=\omega_0-\Omega$ is the detuning between the DQD and the resonator. 
We demonstrate that for DQD with weak energy and phase relaxations, this resonant admittance of the  microwave source to the resonator results in the peak structure of the average photon number and the photocurrent.

In Fig.~\ref{fig:3}, we plot the average number of photons in the resonator and the photocurrent as a function of the drive frequency $\omega$ for $\omega_0=\Omega$ and for the choice of other system parameters identical to those for curves in Fig.~\ref{fig:2}.  At vanishing dephasing rate, $\gamma_\phi=0$, we observe a double peak feature in both photon number and photocurrent curves, see Fig.~\ref{fig:3}. These peaks at $\omega_\pm=(E_{1,\pm}-E_{0})/\hbar$ are defined by the level spacing of the Jaynes-Cummings Hamiltonian and are shown by vertical dashed lines in Fig.~\ref{fig:3}.  The two peaks merge at $\omega=\omega_0$ as the dephasing rate increases  and destroys quantum entanglement between photons and DQD states.

At finite detuning between the resonator and the DQD, $\Delta\gtrsim g=2\pi\times48.5\text{ MHz}$, the eigenstates of the system become  dominantly photon states or electron states of the DQD.  As a result, the microwave source increases the number of photon excitations in the resonator when the microwave frequency is in resonance with the transition between the photon--like states, $\omega_{1,-}=(E_{1,-}-E_0)/\hbar$.  But the source has a weak effect at the resonance with the electron-like states at frequency  $\omega_{1,+}=(E_{1,+}-E_0)/\hbar$. We present the corresponding dependence of the photon number and the photocurrent in Fig.~\ref{fig:4} for $\omega_0/2\pi=8\text{ GHz}$, $\Omega/2\pi=8.1\text{ GHz }(\Delta = 100\text{ MHz})$ and other parameters identical to those for in Figs.~\ref{fig:2} and \ref{fig:3}.  We indeed observe one large peak in the photon number near the resonant condition for the dominantly photon state  with energy $E_{1,-}$ while the photon number does not show significant enhancement near the second resonance, corresponding to the transition to the dominantly electronic state with energy $E_{1,+}$.  
The photocurrent still exhibits double peak feature, but the peak corresponding to the photon resonance is higher, when the microwave drive produces a higher  photon population.

\begin{figure}
\begin{center}
\includegraphics[width=0.95\columnwidth]{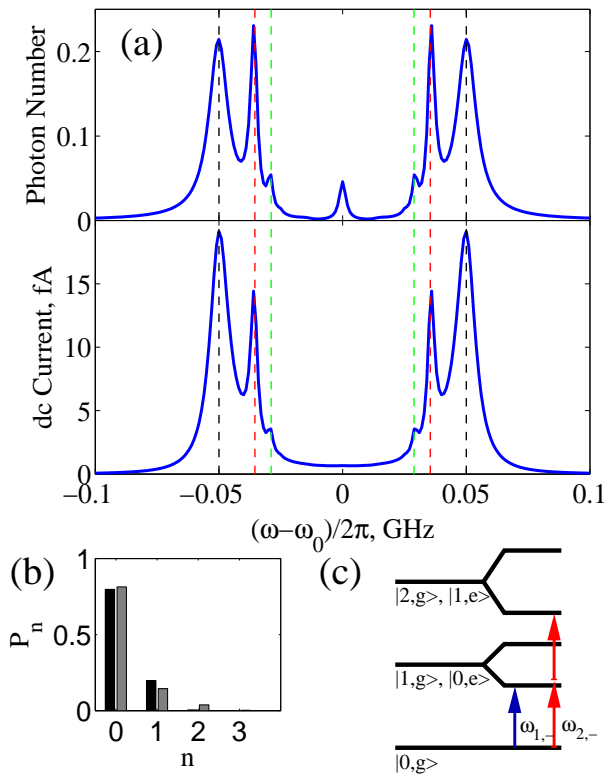}
\par\end{center}
\caption{\label{fig:6}(Color online) (a) The average number of photons in the resonator and the
photovoltaic current as a function of the frequency $\omega$ of the microwave
drive for $\Omega=\omega_{0}=2\pi\times 8\text{ GHz}$,  the intradot tunneling  ${\cal{T}}/2\pi=1$ GHz, and the drive amplitude  ${{F}}/2\pi=30$ MHz and extremely low tunneling rates to the leads and the energy relaxation rate,
$\Gamma_{l}=\Gamma_{r}=\gamma=2\pi\times 100\text{ kHz}$.
The photon average number and the photocurrent have several peaks 
at $\omega_{n,\pm}=(E_{n,\pm}-E_0)/n\hbar$ with $n=1,\, 2,\, 3$, these frequencies, calculated from Eq.~\eqref{eq:JC} are shown by vertical lines). (b) The histogram presents the probabilities $P_n$ to have $n$ photons in the resonator steady state at drive frequency $\omega_{1}$ (dark bars) and $\omega_{2}$ (grey bars).  (c) The diagram is a schematic picture for the Jaynes-Cummings energy levels showing single  and two photon excitations.}
\end{figure}

We now consider a more idealistic regime of significantly reduced tunneling and relaxation rates $\Gamma_{l}=\Gamma_{r}=\gamma=2\pi\times 100\text{ kHz}$,  the drive amplitude $F/2\pi=30$ MHz and $\omega_0=\Omega=2\pi\times 8\text{ GHz}$.  
In this case additional resonances develop, see Fig.~\ref{fig:6}.  
These resonances correspond to excitations of several photons in the cavity by the microwave source.  When the frequency 
of the source satisfies  $\hbar\omega n=E_{n,\pm}-E_0$, the DQD-resonator system experiences transitions from the ground state to the energy state $E_{n,\pm}$, \textit{cf. } Ref.~\onlinecite{Blatt1995}.
These multiphoton transitions result in peaks of the average photon number and the magnitude of the photocurrent. Curves in Fig.~\ref{fig:6} have  three pairs of peaks at frequencies $\omega_{n,\pm}=\omega_0\pm g/\sqrt{n}$ marked by vertical dashed lines for $n=1,2,3$. We notice that for  $\omega=\omega_{1,2}$ the average photon number is nearly the same, see the top panel in Fig.~\ref{fig:6}(a),   
while the photon distribution function is different, Fig.~\ref{fig:6}(b): at $\omega=\omega_{2,-}$ a non-zero $P_2$ develops for a probability that the resonator contains two photons. This difference in  $P_n$ indicates that the microwave drive line does not match the resonator to produce a two photon occupation of the resonator at $\omega=\omega_1$, but it matches the resonator to  populate the state with the energy $E_{2,\pm}$, which then decays to the lower energy states with $n=1,0$.

\begin{figure}
\begin{center}
\includegraphics[width=0.9\columnwidth]{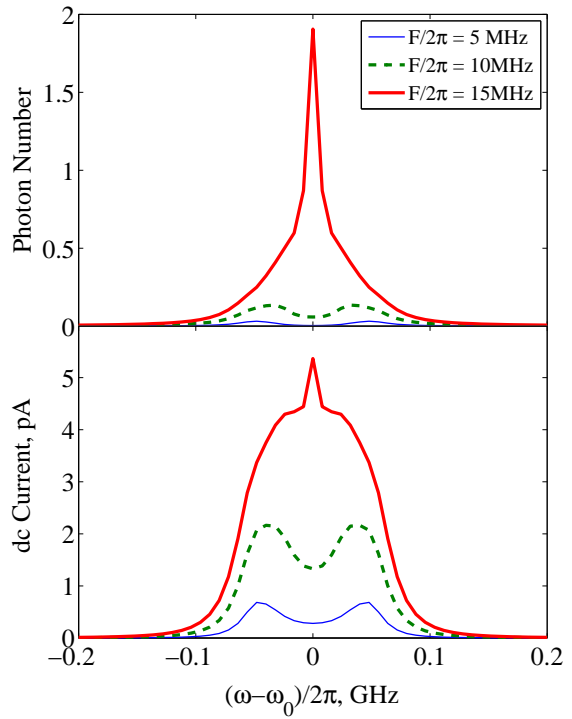}
\par\end{center}
\caption{\label{fig:5}(Color online) Dependence of the average photon number and the current on microwave frequency $\omega$
for several values of the drive amplitude $F/2\pi=5,\, 10,\, 15$ MHz, at zero dephasing $\gamma_\phi=0$ and other parameters are the same as in data in Fig.~\ref{fig:3}. As the amplitude of the drive increases, the two peaks merge together to a single peak at $\omega=\omega_0$.}

\end{figure}

Next, we investigate dependence of the photon number in the resonator
and the magnitude of the photovoltaic current for different amplitudes $F$
of the drive. The above discussion was mostly focused on a resonator
containing less than one photon. As the drive increases, 
the double peak feature evolves to a single peak
at the drive frequency equal to the frequency of the resonator, $\omega=\omega_0$. We
interpret this cross-over as a signature of changed hierarchy of
the terms in the system Hamiltonian. At weak drive, we have a JC Hamiltonian
with its peculiar energy levels, Eq.(\ref{eq:JC}), and the drive can be viewed as a weak
probe testing the spectral structure of the coupled resonator and
DQD system. Once the drive reaches the strength of the $g$ coupling, $g \simeq 2\pi\times 50\text{ MHz}$, a proper way to treat 
the system is to start with the Floquet--type states\cite{Pedersen1998,Marthaler2006,Dykman2012} of the driven resonator and then to take into account the interaction of these states with the DQD system as a perturbation.  In this picture, the photon resonance happens  at $\omega=\omega_0$.  The coupling $g$ is responsible for the formation of the broader ``wings''
in curves for the average photon number and the photocurrent. 
These wings are more pronounced in the photovoltaic current,
which is entirely due to the coupling between resonator and DQD.  This broad
structure of the generated current as a function of the source frequency is preserved even at stronger drive.
Thus, the shape of the photovoltaic curve might provide an experimental approach to quantify
the strength of the JC coupling constant.

\section{Discussion and Conclusion}

We analyzed the photovoltaic current through a DQD system at zero voltage bias between the leads.  The double quantum dot interacts through its  dipole moment to a quantized electromagnetic field of a high quality microwave resonator.  The interaction is described by the Jaynes--Cummings Hamiltonian of a quantized electromagnetic field and a two level quantum system, represented by ground and excited electronic states of the double quantum dot.  When a weak microwave radiation is applied to the resonator, the source acts as a spectral probe that causes excitation of the system when the energy difference between its eigenstates is equal to the photon energy $\hbar\omega$ of the source.  If this resonance condition is satisfied, the microwave source populates the photon mode of the resonator and generates a direct current though the double dot system even at zero bias.  

We demonstrated that at finite, but still low energy and phase relaxation rates of the DQD, both the average number of photons in the resonator and the photocurrent through the DQD have a double-peak structure as functions of the frequency of the microwave source. 
This double peak structure reflects an avoided crossing of the energy states of the DQD and the resonator photons due to the interaction between the two subsystems and is reminiscent of the Lamb shift by a single electromagnetic mode.
We also found that in the limit if extremely weak relaxation rates of the DQD, multiphoton resonances develop when the  energy difference between the states of the coupled system is a multiple of $\hbar\omega$.

As energy and phase relaxation rates of the DQD increase, the peaks in the photon number and the photocurrent 
broaden  and eventually merge in a single resonance peak at the frequency $\omega_0$ of the resonator. In this limit, the resonator photon mode and the DQD are no longer described as an entangled quantum system and the resonant condition for the interaction of the microwave source with the system corresponds to equal frequencies of the source and the resonator mode, $\omega=\omega_0$.

At stronger microwave drive, frequency dependence of the average photon number in the resonator evolve from the Jaynes-Cummings double peaks at $\omega=\omega_0\pm g$ to a single peak at the resonator frequency $\omega_0$.  The single peak at $\omega=\omega_0$ is a result of multi-photon transitions at strong drive by the microwave source that all merge together due to finite width of multi-photon resonances.
Similar evolution to a single peak occurs for the photocurrent response, although the photocurrent curve has a broader width as a function of the source frequency $\omega$, this width corresponds to the strength of the coupling $g$ between the photon mode of the resonator and the DQD and may be used to characterize the strength of this coupling in experiments.

\acknowledgements
We thank R. McDermott and J. Petta for fruitful discussions.  The work was supported by NSF Grant No. DMR-1105178 and by the Donors of the American Chemical Society Petroleum Research Fund.

\end{document}